\documentstyle[12pt]{ioplppt}
\font\cal=cmsy10 scaled 1200
\newcommand{\parsign}{\cal\symbol{'170}}
\newcommand{\Ec}{\cal\symbol{'105}}
\newcommand{\Gc}{\cal\symbol{'107}}
\newcommand{\Hc}{\cal\symbol{'110}}

\newcommand{\Rc}{\cal\symbol{'122}}
\font\semibold=msbm10 scaled 1200
\newcommand{\RB}{\semibold\symbol{'122}}

\begin{document}

\title{Symmetries of Differential Equations via Cartan's
       Method of Equivalence}

\author{O. I. Morozov}

\begin{abstract}
We formulate a method of computing invariant 1-forms and structure equations
of symmetry pseudo-groups of differential equations based on Cartan's method
of equivalence and the moving coframe method introduced by Fels and Olver.
Our apparoach does not require a preliminary computation of infinitesimal
defining systems, their analysis and integration, and uses differentiation
and linear algebra operations only. Examples of its applications are given.
\end{abstract}

\section{Introduction}
The theory of symmetries of differential equations was created by Sophus Lie
more than a hundred years ago.  One of Lie's greatest contributions was
the discovery of the connection between continuous transformation groups and
their infinitesimal ge\-ne\-ra\-tors, which allows one to reduce complicated
nonlinear invariance conditions of d.e.s under  an action of a transformation
group to much simple linear conditions of infinitesimal invariance --
defining equations of symmetry algebra. Lie's method turned out to be a
powerful tool for studying differential equations, finding their exact
solutions, conservation laws, etc.
\cite{Lie,AI,BK,Ibragimov,Ovs,Olver86,Stephani,VKL,VK99}.
In almost all cases the infinitesimal defining equations of the Lie
pseudo-groups of symmetries of d.e.s can be derived algorithmically.
Lie's method requires an integration of (over-determined) system of partial
differential equations to find a symmetry group of d.e.s explicitly.
In the last decade methods  which do not use an integration but rather extract
information about structure of symmetry groups directly from their
infinitesimal defining systems  were developed
\cite{Reid1,Reid2,Schwarz1,Schwarz2}. It was shown how to obtain the dimension
of the finite Lie group, and in \cite{Reid1,Reid2} it was also shown how to
find the structure constants $c^i_{jk}$ of the symmetry algebra in the
finite-dimensional case. In \cite{LisleReidBoulton,LisleReid} the method of
\cite{Reid1,Reid2}  was generalized to the case of structurally transitive
infinite Lie pseudo-groups. Specifically, it was shown how to obtain the
Cartan structure equations of the symmetry pseudo-group  for a system of
d.e.s from its infinitesimal defining equations.

The theory of infinite Lie pseudo-groups was created by {\'E}lie Cartan
\cite{Cartan1} -- \cite{Cartan5}. It does not use infinitesimal methods
and is based on the possibility to characterize an infinite Lie pseudo-group
on a manifold $M$ as the set of projections of bundle transformations
of a principal fiber bundle
$M \times {\mbox{{\Gc}}} \rightarrow M$, where {\Gc} is some Lie group,
that preserve a collection of 1-forms $\tau^i$ on $M \times {\mbox{{\Gc}}}$.
The equations that express the differentials $d \tau^i$  through the
$\tau^i$ and modified Maurer - Cartan forms $\mu^{\alpha}$ of the group {\Gc},
\[
d\tau^i = A^i_{\alpha j}\,\mu^{\alpha}\wedge \tau^j
+ T^i_{jk}\,\tau^j \wedge \tau^k,
\]
\noindent
are called Cartan structure equations; they include important information
about the pseudo-group (see, particularly, \cite[Theorem 11.16]{Olver95}).

In the present paper we apply Cartan's method of equivalence
\cite{Cartan5,Olver95}  and the moving coframe method of \cite{FO,FO2}
to obtain invariant 1-forms of a symmetry pseudo-group of d.e.s.
Unlike the approach of \cite{LisleReidBoulton,LisleReid}, the method used
here does not require a preliminary computation of infinitesimal defining
systems and their reduction to the involutive form.

A system ${\mbox{{\Rc}}}_s$ of differential equations of order $s$ in $n$
independent variables and $m$ dependent variables is locally considered
to be the subbundle in the bundle $J^s(\mbox{{\Ec}})$ of s-jets of the
bundle $\mbox{{\Ec}} =
\mbox{{\RB}}^n \times \mbox{{\RB}}^m \rightarrow \mbox{{\RB}}^n$.
A pseudo-group of symmetries
$Sym({\mbox{{\Rc}}}_s)$ of the system ${\mbox{{\Rc}}}_s$
is a subgroup of the pseudo-group of contact transformations of the bundle
$J^s(\mbox{{\Ec}})$ and consists of those transformations which preserve the
subbundle ${\mbox{{\Rc}}}_s$. So the problem of finding the group
$Sym({\mbox{{\Rc}}}_s)$ is a particular case of the general problem of
equivalence of embedded submanifolds under an action of a pseudo-group.
A powerful and convenient moving coframe method for solving this equivalence
problem was developed in \cite{FO,FO2}.

Some simplifications are possible if we deal with the first order systems of
d.e.s. By \cite[Theorem 3.3.1.]{Pommaret} a system ${\mbox{{\Rc}}}_s$ is
equivalent to the system $\hat{\mbox{{\Rc}}}_1$ of the first order, which is
the  subbundle in $J^1(\hat{\mbox{{\Ec}}})$, where
$\hat{\mbox{{\Ec}}} = J^{s-1}(\mbox{{\Ec}})$. The pseudo-group
$Sym(\hat{\mbox{{\Rc}}}_1)$ of symmetries of the system
$\hat{\mbox{{\Rc}}}_1$ is a subgroup of the pseudo-group
$Cont(J^1(\hat{\mbox{{\Ec}}}))$ of contact transformations of the bundle
$J^1(\hat{\mbox{{\Ec}}})$. By B\"acklund's theorem, \cite{Backlund},
\cite[Theorem 4.32]{Olver95}, contact transformations on
$J^1(\hat{\mbox{{\Ec}}})$ are prolongations of point transformations on
$\hat{\mbox{{\Ec}}}$. Cartan's method of equivalence allows us to obtain
invariant 1-forms which define the pseudo-group of contact transformations.
Then we can find the invariant 1-forms of the pseudo-group
$Sym(\hat{\mbox{{\Rc}}}_1)$. To do that, we should make the following steps.
First, we restrict the invariant 1-forms of the pseudo-group
$Cont(J^1(\hat{\mbox{{\Ec}}}))$ on the subbundle $\hat{\mbox{{\Rc}}}_1$
and obtain the set of linear dependent 1-forms. Next, we apply the procedure
of normalization to the appearing conditions of linear dependence. Finally,
we apply the operations of Cartan's equivalence method to the restrictions on
$\hat{\mbox{{\Rc}}}_1$ of the structure equations of the pseudo-group
$Cont(J^1(\hat{\mbox{{\Ec}}}))$.

\newpage 

\section{Invariant 1-forms and structure equations of the pseudo-group of
contact transformations}

According to
\cite[Theorem 3.3.1.]{Pommaret} a compatible system ${\mbox{{\Rc}}}_s$ of
d.e.s of order $s$ is equivalent to the system
$\hat{\mbox{{\Rc}}}_1$ of order 1, which has more dependent variables.
So it is possible to restrict our attention to the case of $s=1$.

Let ${\mbox{{\Rc}}}_1$ be a system of partial differential equations of the
first order, considered to be the subbundle in the bundle
$J^1(\mbox{{\Ec}})$ of 1-jets of the bundle ${\mbox{{\Ec}}} \rightarrow X$
over an $n$-dimensional base manifold $X$, with $q$-dimensional fibers.
Let $(x^1,x^2,...,x^n)$ denote local coordinates of the base $X$ and
$(u^1,u^2,...,u^q)$ denote local coordinates of the fibers of {\Ec}.
Then local coordinates of the bundle $J^1(\mbox{{\Ec}})$ are
$(x^1,...,x^n,u^1,...,u^q,p^1_1,...,p^1_n,...,p^q_1,...,p^q_n)$, and a
local section $f: X \rightarrow {\mbox{{\Ec}}}$
defined by the equalities $u^\alpha = f^\alpha(x)$,~ $\alpha\in\{1,...,q\}$,~
has corresponding 1-jet $j_1(f) : X \rightarrow J^1({\mbox{{\Ec}}})$, defined
by the equalities $u^\alpha = f^\alpha(x)$,~
$p^\alpha_i={{\partial f^\alpha(x)}\over{\partial x^i}}$,~
$\alpha \in \{1,...,q\}$,~
$i \in \{1,...,n\}$.

A differential form $\tau$ on $J^1(\mbox{{\Ec}})$ is called a {\it contact
form} if it is annihilated by all 1-jets: $j_1(f)^{*}\tau = 0$. In local
coordinates every contact 1-form is a linear combination of the Cartan forms
$\tau^\alpha = d u^\alpha - p^\alpha_i\,d x^i$,~
$\alpha \in \{1,...,q\}$~ (here and below
we use Einstein summation convention, so
$p^\alpha_i\,d x^i = \sum \limits_{i=1}^{n} p^\alpha_i\,d x^i$ etc.)

A local diffeomorphism
$\Delta : J^1(\mbox{{\Ec}}) \rightarrow J^1(\mbox{{\Ec}})$,~
$\Delta : (x,u,p) \mapsto (\overline x,\overline u, \overline p)$, is called a
{\it contact transformation}, if for every contact form $\tau$, the form
$\Delta^{*} \tau$ is also a contact form; in other words, if
 $\Delta^{*}\tau^\alpha =
d {\overline u}^\alpha - {\overline p}^\alpha_i \, d {\overline x}^i =
\zeta^\alpha_\beta(x,u,p)\,\tau^\beta$ for some functions $\zeta^\alpha_\beta$
on $J^1(\mbox{{\Ec}})$.

By B\"acklund's theorem, \cite{Backlund}, \cite[Theorem 4.32]{Olver95},
in the case of
$n>1$ and $q>1$ every contact transformation
$\Delta : J^1(\mbox{{\Ec}}) \rightarrow J^1(\mbox{{\Ec}})$
is a prolongation of a point transformation
$\Gamma : \mbox{{\Ec}} \rightarrow \mbox{{\Ec}}$,~
$\Gamma : (x,u) \mapsto (\overline x, \overline u)$, where the functions
$\overline p^\alpha_i$ are defined by the equalities
\begin{equation}
{{\partial {\overline u}^\alpha}\over{\partial x^j}}+
{{\partial {\overline u}^\alpha}\over{\partial u^\beta}}\,p^\beta_j =
\overline p^\alpha_i\,\left(
{{\partial {\overline x}^i}\over{\partial x^j}}+
{{\partial {\overline x}^i}\over{\partial u^\beta}}\,p^\beta_j
\right).
\label{p_rules}
\end{equation}

To obtain a collection of invariant 1-forms of the pseudo-group of contact
trans\-for\-ma\-ti\-ons on
$J^1(\mbox{{\Ec}})$ we apply Cartan's equivalence method
\cite{Cartan5,Olver95}. For this purpose we consider the coframe
$\left\{(\tau^\alpha, d x^i, d p^\alpha_i)\, \vert \,\alpha \in \{1,...,q\},
i\in\{1,...,n\}\,\right\}$ on $J^1(\mbox{{\Ec}}).$
A contact transformation $\Delta$ acts on this coframe in the following
manner:
\[
\Delta^{*}\,\left(\begin{array}{c}\tau^\alpha\\
d x^i\\d p^\alpha_i\end{array}\right)=
S\,\left(\begin{array}{c}\tau^\alpha\\d x^i\\d p^\alpha_i\end{array}\right),
\]
\noindent where $S : J^1({\mbox{{\Ec}}}) \rightarrow {\mbox{{\Gc}}}$ is
an analytic function on $J^1({\mbox{{\Ec}}})$ taking values in the Lie group
${\mbox{{\Gc}}}$ of non-degenerate block lower triangular matrices of the form
\[
\left(
\begin{array}{ccc}
a^\alpha_\beta & 0 & 0\\
C^i_\beta & b^i_j & 0\\
F^\alpha_{i\beta}&G^\alpha_{ij}&h^{\alpha j}_{i \beta}
\end{array}
\right).
\]

In accordance with Cartan's method of equivalence, we consider the lifted
coframe on
$J^1({\mbox{{\Ec}}})\times {\mbox{{\Gc}}}$
\begin{eqnarray}
\Theta^\alpha = a^\alpha_\beta\, \tau^\beta,\nonumber\\
\Xi^i = c^i_\beta\, \Theta^\beta +b^i_j\, d x^j,  \label{lcf}\\
\Sigma^\alpha_i = f^\alpha_{i\beta}\,\Theta^\beta+g^\alpha_{ij}\, \Xi^j +
h^{\alpha j}_{i\beta}\,d p^\beta_j, \nonumber
\end{eqnarray}
\noindent
where for convenience we use the notations
$c^i_\beta=C^i_\gamma\,A^\gamma_\beta$,~
$f^\alpha_{i \beta} = F^\alpha_{i \gamma}\,A^\gamma_\beta
- G^\alpha_{ij}\,B^j_k\,c^k_\beta$~,
$g^\alpha_{ij}=G^\alpha_{ik}\,B^k_j$~;
 $(A^\beta_\gamma)$ is the inverse matrix of the matrix $(a^\alpha_\beta)$,
$(B^j_k)$ is the inverse matrix of the matrix  $(b^i_j)$,
so $a^\alpha_\beta\,A^\beta_\gamma = \delta^\alpha_\gamma$
and $b^i_j\,B^j_k = \delta^i_k$.
To find an invariant coframe we use the procedure of absorption and
normalization of essential torsion coefficients
\cite[Chapter 10]{Olver95}.

Taking exterior differentials of 1-forms $\Theta^\alpha$ and
substituting the differentials $du^\beta$, $d x^j$, $d p^\beta_j$
expressed from the equations (\ref{lcf}), we obtain
\begin{eqnarray*}
d \Theta^\alpha = \left(d a^\alpha_\beta\,A^\beta_\gamma
+a^\alpha_\beta\,B^j_k\,H^{\beta s}_{j \eta} \left(
c^k_\gamma\,(\Sigma^\eta_s-f^\eta_{s\epsilon}\,\Theta^\epsilon
-g^\eta_{sl}\,\Xi^l)-f^\eta_{s\gamma}\,\Xi^s
\right)\right)\wedge\Theta^\gamma \\~~~~~~~~~
+a^\alpha_\beta\,B^j_k\,H^{\beta s}_{j\eta}\, \Xi^k\wedge \Sigma^\eta_s
-a^\alpha_\beta\,B^j_k\,H^{\beta s}_{j\eta}\,g^\eta_{sl}\,
\Xi^k\wedge \Xi^l,
\end{eqnarray*}
\noindent
where the functions $H^{\beta k}_{j \gamma}$ are defined by the conditions
$H^{\beta k}_{j \gamma} \,h^{\gamma i}_{k \alpha} =
\delta^i_j\,\delta^\beta_\alpha$.
The multipliers of $\Xi^k\wedge \Sigma^\eta_s$ and $\Xi^k\wedge \Xi^l$ are
essential torsion coefficients. We normalize them by the
following choice of the parameters of the Lie group {\Gc}:
\begin{eqnarray}
h^{\alpha j}_{i \beta} = a^\alpha_\beta\,B^j_i, \label{def_h}\\
g^\alpha_{ij}=g^\alpha_{ji}. \label{symm_g}
\end{eqnarray}
\noindent Then we have
\begin{eqnarray}
d \Theta^\alpha = \Phi^\alpha_\beta \wedge \Theta^\beta
+ \Xi^k \wedge \Sigma^\alpha_k,\label{se1}\\
\Phi^\alpha_\beta=d a^\alpha_\gamma\,A^\gamma_\beta
+c^k_\gamma\,f^\alpha_{k\beta}\,\Theta^\gamma-f^\alpha_{k\beta}\,\Xi^k
-c^k_\beta\,g^\alpha_{kj}\,\Xi^j+c^k_\beta\,\Sigma^\alpha_k. \label{pcf1}
\end{eqnarray}
\noindent
Now the exterior differentials of $\Xi^i$ and $\Sigma^\alpha_i$ become
\begin{eqnarray}
d \Xi^i = \Psi^i_k\wedge \Xi^k +\Pi^i_\gamma \wedge \Theta^\gamma,
\label{se2}\\
d \Sigma^\alpha_i = \Phi^\alpha_\gamma\wedge\Sigma^\gamma_i
-\Psi^k_i\wedge\Sigma^\alpha_k+\Lambda^\alpha_{i\beta}\wedge\Theta^\beta
+\Omega^\alpha_{ij}\wedge\Xi^j, \label{se3}
\end{eqnarray}
\noindent
where
\begin{eqnarray}
\Psi^i_k = d b^i_j\,B^j_k - c^i_\beta\,\Sigma^\beta_k, \label{pcf2}\\
\Pi^i_\gamma = d c^i_\gamma+c^i_\beta\,\Phi^\beta_\gamma-
c^k_\gamma\,\Psi^i_k-c^k_\gamma\,c^i_\beta\,\Sigma^\beta_k, \label{pcf3}\\
\Lambda^\alpha_{i\beta} = d f^\alpha_{i\beta}
+f^\alpha_{i\gamma}\,\Phi^\gamma_\beta+g^\alpha_{ij}\,\Pi^j_\beta
-f^\gamma_{i\beta}\,(\Phi^\alpha_\gamma
-c^k_\epsilon\,f^\alpha_{k\gamma}\,\Theta^\epsilon+f^\alpha_{k\gamma}\,\Xi^k
\nonumber\\
~~~~~~~+c^k_\gamma\,g^\alpha_{kj}\,\Xi^j-c^k_\gamma\,\Sigma^\alpha_k)
+f^\alpha_{k\beta}\,(\Psi^k_i+c^k_\gamma\,\Sigma^\gamma_i)
+c^k_\beta\,f^\alpha_{k\gamma}\,\Sigma^\gamma_i, \label{pcf4}\\
\Omega^\alpha_{ij} = d g^\alpha_{ij}+g^\alpha_{ik}\,\Psi^k_j
+g^\alpha_{jk}\,\Psi^k_i-f^\alpha_{i\beta}\,\Sigma^\beta_j
-f^\alpha_{j\beta}\,\Sigma^\beta_i\nonumber\\
~~~~~~~-g^\gamma_{ij}\,(\Phi^\alpha_\gamma
-c^k_\beta\,f^\alpha_{k\gamma}\,\Theta^\beta+f^\alpha_{k\gamma}\Xi^k
+c^k_\gamma\,g^\alpha_{ks}\,\Xi^s-c^k_\gamma\,\Sigma^\alpha_k). \label{pcf5}
\end{eqnarray}
\noindent
We note that the conditions (\ref{symm_g}) imply
\begin{equation}
\Omega^\alpha_{ij}=\Omega^\alpha_{ji}. \label{symm_mu}
\end{equation}
Thus the specifications (\ref{def_h}) and (\ref{symm_g})
of the group parameters of the coframe (\ref{lcf}) give the lifted coframe
\begin{eqnarray}
\Theta^\alpha = a^\alpha_\beta\,(d u^\beta - p^\beta_j\,d x^j),
\label{theta}\\
\Xi^i= c^i_\beta\,\Theta^\beta+b^i_j\,d x^j, \label{xi}\\
\Sigma^\alpha_i= f^\alpha_{i\beta}\,\Theta^\beta+g^\alpha_{ij}\,\Xi^j
+a^\alpha_\beta\,B^j_i\,d p^\beta_j  \label{tau}
\end{eqnarray}
\noindent
on  $J^1({\mbox{{\Ec}}})\times {\mbox{{\Hc}}}$, where {\Hc} is the subgroup
of the group {\Gc} defined by the equalities (\ref{def_h}) and (\ref{symm_g}).
The structure equations (\ref{se2}), (\ref{se3}) do not contain any torsion
coefficients, while the structure equations (\ref{se1}) contain only constant
torsion coefficients.

The structure equations (\ref{se1}), (\ref{se2}), (\ref{se3}) remain unchanged
if we make the following change of the modified Maurer - Cartan forms
$\Phi^\alpha_\beta$, $\Psi^i_k$, $\Pi^i_\gamma$, $\Lambda^\alpha_{i\beta}$,
$\Omega^\alpha_{ij}$~ :
\[
\begin{array}{lll}
\Phi^\alpha_\beta &\mapsto& \Phi^\alpha_\beta
+ K^\alpha_{\gamma\epsilon}\,\Theta^\epsilon,\\
\Psi^i_k &\mapsto& \Psi^i_k + L^i_{kj}\,\Xi^j+M^i_{k\gamma}\,\Theta^\gamma,\\
\Pi^i_\gamma &\mapsto& \Pi^i_\gamma + M^i_{k\gamma}\,\Xi^k
+N^i_{\gamma \epsilon}\,\Theta^\epsilon,\\
\Lambda^\alpha_{i\beta} &\mapsto& \Lambda^\alpha_{i\beta} +
P^\alpha_{i\beta\gamma}\,\Theta^\gamma+Q^\alpha_{i\beta k}\,\Xi^k
+ K^\alpha_{\gamma\beta}\,\Sigma^\gamma_i-M^k_{i\beta}\,\Sigma^\alpha_k,\\
\Omega^\alpha_{ij} &\mapsto& \Omega^\alpha_{ij}
+ Q^\alpha_{i\beta j}\,\Theta^\beta
+R^\alpha_{ijk}\,\Xi^k - L^k_{ij}\,\Sigma^\alpha_k,
\end{array}
\]
\noindent
where $K^\alpha_{\gamma\epsilon}$, $L^i_{kj}$, $M^i_{k\gamma}$,
$N^i_{\gamma \epsilon}$, $P^\alpha_{i\beta\gamma}$, $Q^\alpha_{i\beta k}$,
$R^\alpha_{ijk}$ are arbitrary functions on
$J^1(\mbox{{\Ec}})\times {\mbox{{\Hc}}}$  satisfying the following symmetry
conditions:
\begin{eqnarray}
K^\alpha_{\gamma\epsilon}=K^\alpha_{\epsilon\gamma},~~
L^i_{kj}=L^i_{jk},~~
N^i_{\gamma \epsilon}=N^i_{\epsilon\gamma}, \nonumber\\
P^\alpha_{i\beta\gamma}=P^\alpha_{i\gamma\beta},~~
Q^\alpha_{i\beta k}=Q^\alpha_{k\beta i},~~
R^\alpha_{ijk}=R^\alpha_{ikj}=R^\alpha_{jik}.\label{sym_cond}
\end{eqnarray}
\noindent
Their number
\begin{eqnarray*}
r^{(1)} =\case12\,q^2\,(q+1)+\case12\,n^2\,(n+1) +n^2\,q
+\case12\,n\,q\,(q+1) + \case12\,n\,q^2\,(q+1)\\
~~~~~+\case12\,n\,q^2\,(n+1)
+\case16\,q\,n\,(n+1)\,(n+2)
\end{eqnarray*}
\noindent
is the degree of indeterminancy \cite[Definition 11.2]{Olver95}
of the lifted coframe $\Theta^\alpha$, $\Xi^i$, $\Sigma^\alpha_i$.

Using the conditions (\ref{symm_mu}), it is not hard to compute the reduced
characters \cite[Definition 11.4]{Olver95} of this coframe:
$s^{\prime}_1 = s^{\prime}_2 = ... = s^{\prime}_q = q+n+n\,q$,
$s^{\prime}_{q+1} = n+n\,q$,
$s^{\prime}_{q+2} = n+(n-1)\,q$,
$s^{\prime}_{q+2} = n+(n-2)\,q$, ... ,
$s^{\prime}_{q+n-1} = n+2\,q$,
$s^{\prime}_{q+n} = n+q$,
$s^{\prime}_{q+n+1} = s^{\prime}_{q+n+2} = ... =
s^{\prime}_{q+n+nq} = 0$.
It is easy to verify that the Cartan test
\[
r^{(1)} =s^{\prime}_1+2\,s^{\prime}_2+3\,s^{\prime}_3 + ...
+ (q+n+n\,q)\,s^{\prime}_{q+n+nq}
\]
\noindent
is satisfied, so by definition  11.7 of \cite{Olver95}
the lifted coframe (\ref{theta}), (\ref{xi}), (\ref{tau}) is involutive, and
by theorem 11.16 of \cite{Olver95}, since the last non-zero reduced
character $s^{\prime}_{q+n}$ is equal to $q+n$,
the transformations of the invariance pseudo-group of this coframe depend on
$q+n$ functions of $q+n$ variables, as it should be. It is easy to verify
directly that the transformation
$\Upsilon : J^1(\mbox{{\Ec}})\times {\mbox{{\Hc}}}  \rightarrow
J^1(\mbox{{\Ec}})\times {\mbox{{\Hc}}}$
satisfies the conditions
\begin{equation}
\Upsilon^{*}\,\Theta^\alpha=\Theta^\alpha, ~~~
\Upsilon^{*}\,\Xi^i=\Xi^i, ~~~
\Upsilon^{*}\,\Sigma^\alpha_i=\Sigma^\alpha_i \label{contact_invariant}
\end{equation}
\noindent if and only if it is projectable on $J^1(\mbox{{\Ec}})$
and its projection
$\Delta : J^1(\mbox{{\Ec}}) \rightarrow J^1(\mbox{{\Ec}})$,
$\Delta : (x,u,p) \mapsto (\overline x, \overline u, \overline p)$,
is the prolongation of the transformation
$\Gamma : \mbox{{\Ec}} \rightarrow \mbox{{\Ec}}$,
$\Gamma : (x,u) \mapsto (\overline x, \overline u)$,
such that the conditions
(\ref{p_rules}) are satisfied.  Thus the equalities (\ref{contact_invariant})
really define the pseudo-group of contact transformations on
$J^1(\mbox{{\Ec}})$, when $q>1$ and $n>1$.

Since the forms $\Theta^\alpha$, $\Xi^i$, $\Sigma^\alpha_i$ are preserved
by the pseudo-group transformations, their exterior differentials are
preserved also, so
$\Upsilon^{*} d\Theta^\alpha = d\Theta^\alpha$,
$\Upsilon^{*} d\Xi^i = d\Xi^i$,
$\Upsilon^{*} d\Sigma^\alpha_i = d\Sigma^\alpha_i$,
therefore we have
\begin{eqnarray*}
\fl
\Upsilon^{*}(\Phi^\alpha_\beta \wedge \Theta^\beta
+ \Xi^k \wedge \Sigma^\alpha_k)
=(\Upsilon^{*}\Phi^\alpha_\beta) \wedge \Theta^\beta
+ \Xi^k \wedge \Sigma^\alpha_k=
\Phi^\alpha_\beta \wedge \Theta^\beta + \Xi^k \wedge \Sigma^\alpha_k,\\
\fl
\Upsilon^{*}(\Psi^i_k\wedge \Xi^k +\Pi^i_\gamma \wedge \Theta^\gamma)=
(\Upsilon^{*}\Psi^i_k)\wedge \Xi^k
+(\Upsilon^{*}\Pi^i_\gamma) \wedge \Theta^\gamma=
\Psi^i_k\wedge \Xi^k +\Pi^i_\gamma \wedge \Theta^\gamma,\\
\fl
\Upsilon^{*}(\Phi^\alpha_\gamma\wedge\Sigma^\gamma_i
-\Psi^k_i\wedge\Sigma^\alpha_k
+\Lambda^\alpha_{i\beta}\wedge\Theta^\beta+\Omega^\alpha_{ij}\wedge\Xi^j)\\
=\Upsilon^{*}(\Phi^\alpha_\gamma)\wedge\Sigma^\gamma_i
-(\Upsilon^{*}\Psi^k_i)\wedge\Sigma^\alpha_k
+(\Upsilon^{*}\Lambda^\alpha_{i\beta})\wedge\Theta^\beta
+(\Upsilon^{*}\Omega^\alpha_{ij})\wedge\Xi^j\\
=\Phi^\alpha_\gamma\wedge\Sigma^\gamma_i-\Psi^k_i\wedge\Sigma^\alpha_k
+\Lambda^\alpha_{i\beta}\wedge\Theta^\beta+\Omega^\alpha_{ij}\wedge\Xi^j,
\end{eqnarray*}
\noindent
and thus
\begin{eqnarray}
\Upsilon^{*}\Phi^\alpha_\beta = \Phi^\alpha_\beta
+ K^\alpha_{\gamma\epsilon}\,\Theta^\epsilon,\nonumber\\
\Upsilon^{*}\Psi^i_k = \Psi^i_k + L^i_{kj}\,\Xi^j
+M^i_{k\gamma}\,\Theta^\gamma,\nonumber\\
\Upsilon^{*}\Pi^i_\gamma = \Pi^i_\gamma + M^i_{k\gamma}\,\Xi^k
+N^i_{\gamma \epsilon}\,\Theta^\epsilon,\label{trans_pcf}\\
\Upsilon^{*}\Lambda^\alpha_{i\beta} = \Lambda^\alpha_{i\beta} +
P^\alpha_{i\beta\gamma}\,\Theta^\gamma+Q^\alpha_{i\beta k}\,\Xi^k
+ K^\alpha_{\gamma\beta}\,\Sigma^\gamma_i
-M^k_{i\beta}\,\Sigma^\alpha_k,\nonumber\\
\Upsilon^{*}\Omega^\alpha_{ij} = \Omega^\alpha_{ij}
+ Q^\alpha_{i\beta j}\,\Theta^\beta
+R^\alpha_{ijk}\,\Xi^k - L^k_{ij}\,\Sigma^\alpha_k\nonumber
\end{eqnarray}
\noindent
for some functions
$K^\alpha_{\gamma\epsilon}$, $L^i_{kj}$, $M^i_{k\gamma}$,
$N^i_{\gamma \epsilon}$, $P^\alpha_{i\beta\gamma}$, $Q^\alpha_{i\beta k}$,
$R^\alpha_{ijk}$ on $J^1(\mbox{{\Ec}})\times {\mbox{{\Hc}}}$
satisfying the conditions (\ref{sym_cond}).

\section{Symmetries of differential equations}

A suitable method for studying geometrical properties of embedded submanifolds
under an action of finite-dimensional Lie groups or infinite Lie pseudo-groups
was developed in \cite{FO,FO2}. For its application to the problem of finding
symmetries of a system of d.e.s ${\mbox{{\Rc}}}_1$ we restrict the lifted
coframe (\ref{theta}), (\ref{xi}), (\ref{tau}) on ${\mbox{{\Rc}}}_1$. That is,
we consider the set of 1-forms $\theta^\alpha = \iota^{*} \Theta^\alpha$,
$\xi^i = \iota^{*} \Xi^i$, $\sigma^\alpha_i = \iota^{*} \Sigma^\alpha_i$,
where $\iota : {\mbox{{\Rc}}}_1 \rightarrow J^1({\mbox{{\Ec}}})$ is the
embedding (for brevity we identify the map
$\iota\times id : {\mbox{{\Rc}}}_1 \times {\mbox{{\Hc}}}
\rightarrow J^1({\mbox{{\Ec}}})\times {\mbox{{\Hc}}}$
with $\iota : {\mbox{{\Rc}}}_1 \rightarrow J^1({\mbox{{\Ec}}})$).
The 1-forms $\theta^\alpha$, $\xi^i$, $\sigma^\alpha_i$ are linearly
dependent, i.e., there exists a non-trivial set of functions $U_\alpha$,
$V_i$, $W^i_\alpha$ on ${\mbox{{\Rc}}}_1 \times {\mbox{{\Hc}}}$, such that
$U_\alpha\,\theta^\alpha+V_i\,\xi^i+W^i_\alpha\,\sigma^\alpha_i\equiv 0$.

Setting these functions equal to some constants allows one to express a part
of parameters $a^\alpha_\beta$, $b^i_j$, $c^i_\beta$, $f^\alpha_{i\beta}$,
$g^\alpha_{ij}$ of the group {\Hc} as functions of coordinates
of ${\mbox{{\Rc}}}_1$ and other group parameters. Substituting the obtained
values of parameters into the modified Maurer - Cartan forms
$\phi^\alpha_\beta=\iota^{*}\Phi^\alpha_\beta$,
$\psi^i_k=\iota^{*}\Psi^i_k$, $\pi^i_\beta=\iota^{*}\Pi^i_\beta$,
$\lambda^\alpha_{i\beta}=\iota^{*}\Lambda^\alpha_{i\beta}$,
$\omega^\alpha_{i j}=\iota^{*}\Omega^\alpha_{i j}$
makes a part of these forms independent of all differentials of the group
parameters. Since the transformation $\Upsilon^{*}$ changes the forms
$\Phi^\alpha_\beta$, $\Psi^i_k$, $\Pi^i_\beta$ by the rules
(\ref{trans_pcf}), in the case when the obtained form $\phi^\alpha_\beta$
does not depend on all differentials of the group parameters, its coefficients
at $\sigma^\gamma_j$ and $\xi^j$ are lifted invariants of the pseudo-group,
and if the obtained forms $\psi^i_k$ or $\pi^i_\beta$ are independent of all
differentials of the group parameters, their coefficients at
$\sigma^\gamma_j$ are lifted invariants also. Normalizing these lifted
invariants to be constants allows us to express a part of the group parameters
as functions of coordinates on ${\mbox{{\Rc}}}_1$ and other group parameters.
If not all group parameters are expressed, we should substitute the expressed
parameters into the forms $\phi^\alpha_\beta$, $\psi^i_k$, $\pi^i_\gamma$,
which depend on their differentials, and repeat the process. If the process is
completed, but not all group parameters are expressed as functions on
${\mbox{{\Rc}}}_1$, we should substitute the modified Maurer - Cartan forms
$\phi^\alpha_\beta$, $\psi^i_k$, $\pi^i_\gamma$, $\lambda^\alpha_{i\beta}$,
$\omega^\alpha_{ij}$, which were reduced during the process of normalization,
into the reduced structure equations
\begin{eqnarray*}
d \theta^\alpha = \phi^\alpha_\beta \wedge \theta^\beta
+ \xi^k \wedge \sigma^\alpha_k,\\
d \xi^i = \psi^i_k\wedge \xi^k
+\pi^i_\gamma \wedge \theta^\gamma,\\
d \sigma^\alpha_i = \phi^\alpha_\gamma\wedge\sigma^\gamma_i
-\psi^k_i\wedge\sigma^\alpha_k
+\lambda^\alpha_{i\beta}\wedge\theta^\beta
+\omega^\alpha_{ij}\wedge\xi^j.
\end{eqnarray*}
If the essential torsion coefficients dependent on the group parameters
appear, then we should normalize them to constants and find some new part of
the group parameters, which, being substituted into the reduced modified
Maurer - Cartan forms, allows us to repeat the procedure of normalization.
There are two possible results of this process. The first one, when the
reduced lifted coframe appears to be involutive, outputs the desired set of
invariant 1-forms which characterize the pseudo-group $Sym({\mbox{{\Rc}}}_1)$.
In the second one, when the coframe is not involutive, we should apply the
procedure of prolongation \cite[Chapter 12]{Olver95}.

\subsection{Example 1: Burgers' equation}

For an application of the above method to finding invariant 1-forms of the
symmetry group of the Burgers' equation
\[
u_t=u_{xx}+u\,u_x,
\]
\noindent
we take the equivalent system of the first order
\[
u_x=v,~~~~~
v_x=u_t-u\,v.
\]
\noindent
Denoting $x=x^1$, $t=x^2$, $v=u^1$, $u=u^2$, $v_x=p^1_1$, $v_t=p^1_2$,
$u_x=p^2_1$, $u_t=p^2_2$, we consider this system as a subbundle of
the bundle $J^1({\mbox{{\Ec}}})$,~ ${\mbox{{\Ec}}}={\mbox{{\RB}}}^2
\times {\mbox{{\RB}}}^2 \rightarrow {\mbox{{\RB}}}^2$,~ with local coordinates
$\{x^1,x^2,u^1,u^2,p^1_1,p^1_2,p^2_1,p^2_2\}$, where the embedding $\iota$ is
defined by the equalities
\[
p^1_1=p^2_2-u^1\,u^2,~~~~~ p^2_1=u^1.
\]
\noindent
The forms
$\theta^\alpha = \iota^{*}\,\Theta^\alpha$, $\alpha\in\{1,2\}$,
$\xi^i = \iota^{*}\,\Xi^i$, $i\in\{1,2\}$, are linearly independent,
whereas the forms $\sigma^\alpha_i = \iota^{*}\,\Sigma^\alpha_i$ are
linearly dependent. The group parameters $a^\alpha_\beta$, $b^i_j$ must
satisfy the conditions $det \left(a^\alpha_\beta\right) \not = 0$,
$det \left(b^i_j\right) \not = 0$. Moreover, without loss of generality,
we can consider that $a^1_1\not =0$, $a^2_2\not =0$, $b^1_1\not =0$,
$b^2_2\not =0$. Computing the linear dependence conditions of forms
$\sigma^\alpha_i$ by means of {\sc MAPLE}, we obtain
sequentially the group parameters $a^2_1$, $b^2_1$, $b^2_2$, $g^2_{12}$,
$g^2_{11}$, $g^1_{11}$, $f^2_{12}$, $f^2_{11}$, $g^2_{22}$, $f^2_{22}$,
$f^2_{211}$ as the functions of other group parameters and the local
coordinates $\{x^1,x^2,u^1,u^2,p^1_2,p^2_2\}$ of ${\mbox{{\Rc}}}_1$.
Particularly,
\begin{eqnarray*}\fl
a^2_1=0,~~ b^2_1=0,~~ b^2_2={\frac {b^1_1a^2_2}{a^1_1}},~~
g^2_{12}=-{\frac {\left (-p^2_2b^1_2+u^1u^2b^1_2
+p^1_2b^1_1\right )a^1_1}{(b^1_1)^{3}}},\\
\fl g^2_{11}=-{\frac {a^2_2\left (p^2_2-u^1u^2\right )}
{(b^1_1)^{2}}},~~
g^1_{11}={\frac {(u^1)^{2}a^1_1-a^1_2p^2_2
-a^1_1p^1_2-u^1(u^2)^{2}a^1_1+p^2_2a^1_1u^2
+u^1u^2a^1_2}{(b^1_1)^{2}}},\\
\fl f^2_{12}={\frac {(b^1_1)^{2}a^1_2+p^1_2(a^1_1)^{2}c^2_2
b^1_1+u^1u^2(a^1_1)^{2}b^1_2c^2_2-u^1u^2a^2_2
c^1_2b^1_1a^1_1-p^2_2(a^1_1)^{2}b^1_2c^2_2
+p^2_2a^2_2c^1_2b^1_1a^1_1}{(b^1_1)^{3}a^1_1}},\\
\fl f^2_{11}=-{\frac {u^1u^2(c^1_1a^2_2b^1_1a^1_1-(a^1_1)^{2}b^1_2c^2_1)
-p^2_2c^1_1a^2_2b^1_1a^1_1
+p^2_2(a^1_1)^{2}b^1_2c^2_1
-p^1_2(a^1_1)^{2}c^2_1b^1_1
+a^2_2(b^1_1)^{2}}{(b^1_1)^{3}a^1_1}},\\
\end{eqnarray*}
\noindent
while the expressions for $g^2_{22}$, $f^2_{22}$ and $f^2_{21}$ are too big
to write them out in full
here.

The linear dependence between the forms $\sigma^\alpha_i$ is
$\sigma^1_1=\sigma^2_2$, $\sigma^2_1 = 0$.

The analysis of the modified Maurer - Cartan forms
$\phi^\alpha_\beta$, $\psi^i_k$, $\pi^i_\gamma$ at the obtained values of
the group parameters gives the following normalizations:
\begin{eqnarray*}
\fl
\phi^2_1 \equiv c^2_1\,\sigma^2_2+{{a^2_2}\over{b^1_1 a^1_1}}\,
\xi^1 ~~ \left({\mbox{{\rm mod}}}~~ \theta^1,~\theta^2,~
\xi^2,~ \sigma^1_2\right)
~~\Rightarrow~~
c^2_1 = 0,~~~ b^1_1 = {{a^2_2}\over{a^1_1}};\\
\fl
{\psi}^2_2 - 2\,{\psi}^1_1 =(2\,c^1_1-c^2_2)\,{\sigma}^1_2
~~\Rightarrow~~
c^2_2=2\,c^1_1;\\
\fl
\psi^1_1+\phi^1_1-\phi^2_2 \equiv
-2\,c^1_1 \sigma^2_2
~~\left({\mbox{{\rm mod}}}~~ \theta^1,~\theta^2,~ \xi^1,~
\xi^2,~ \sigma^1_2\right)
~~\Rightarrow~~
c^1_1=0;\\
\fl
\phi^2_1 \equiv
-\left(f^1_{11} + {{(a^1_2a^2_2-a^1_1a^2_2u^2+b^1_2(a^1_1)^2}\over{a^2_2}}
\right)\,
\xi^2
~~\left({\mbox{{\rm mod}}}~~ \theta^1,~\theta^2,~ \xi^1,~
\sigma^1_2, ~ \sigma^2_2\right)\\
~~\Rightarrow~~
f^1_{11}= - {{a^1_2a^2_2-a^1_1a^2_2u^2+b^1_2(a^1_1)^2}\over{a^2_2}}.
\end{eqnarray*}

Now the analysis of the structure equations gives step by step the following
essential torsion coefficients and the corresponding normalizations:
\begin{eqnarray*}
\fl d \theta^1 = -c^1_2\,\theta^2 \wedge \sigma^2_2+...
~~\Rightarrow~~
c^1_2 = 0;\\
\fl d \theta^1 = \left((a^2_2)^3 f^1_{12} - (a^1_2)^2a^2_2+a^1_1a^1_2a^2_2u^2
-(a^1_1)^2a^1_2b^1_2\right)\,\theta^2 \wedge \xi^1
+ \left(f^1_{22}+{{a^1_2}\over{a^2_2}}\, f^1_{21}\right)\,\theta^2
\wedge \xi^2+...\\
~~\Rightarrow~~
f^1_{12}= {{a^1_2 (a^1_2a^2_2-a^1_1a^1_2a^2_2u^2+(a^1_1)^2a^1_2b^1_2)}
\over{(a^2_2)^{3}}},~~~
f^1_{22}= -{{a^1_2}\over{a^2_2}}\, f^1_{21};\\
\fl d \xi^2={{2\,(2\,a^1_2a^2_2-a^1_1a^2_2u^2+b^1_2(a^1_1)^2)}
\over{(a^2_2)^{2}}}\,\xi^1 \wedge \xi^2+...
~~\Rightarrow~~
a^1_2 = {{a^1_1\,(a^2_2u^2-b^1_2a^1_1)}\over{2\,a^2_2}};\\
\fl d \xi^1=\left(f^2_{21} +{{(a^1_1)^2\,(4\,(a^2_2)^2u^1
-2\,a^2_2b^1_2a^1_1u^2+(a^2_2)^2(u^2)^2+(b^1_2)^2(a^1_1)^2)}\over
{(a^2_2)^{4}}}\right)\,\xi^1 \wedge \xi^2+...\\
~~\Rightarrow~~
f^2_{21} = -{{(a^1_1)^2\,(4\,(a^2_2)^2u^1-2\,a^2_2b^1_2a^1_1u^2
+(a^2_2)^2(u^2)^2+(b^1_2)^2(a^1_1)^2)}\over{(a^2_2)^{4}}};\\
\fl
d \sigma^1_2=-{{(a^1_1)^2\,(b^1_2 a^1_1-a^2_2 u^2)}
\over{(a^2_2)^{4}}}\theta^1 \wedge \theta^2+...
~~\Rightarrow~~
b^1_{2} = {{a^2_2u^2}\over{a^1_1}};\\
\fl
d \sigma^2_2={{(a^1_1)^3\,(p^2_2-u^1u^2)}\over{(a^2_2)^{3}}}\theta^2
\wedge \xi^1+...
~~\Rightarrow~~
a^2_2 = {{a^1_1}\over{(p^2_2-u^1u^2)^{1/3}}};\\
\fl d \theta^2={1\over{3\,a^1_1\,(p^2_2-u^1u^2)^{2/3}}}\theta^2
\wedge \sigma^2_2+...
~~\Rightarrow~~
a^1_1 = {1\over{(p^2_2-u^1u^2)^{2/3}}};\\
\fl
d \theta^1=-\left({{2\,g^1_{12}}\over 3} + {{2\,u^1}
\over{(p^2_2-u^1u^2)^{2/3}}}\right)\theta^1 \wedge \xi_2+...
~~\Rightarrow~~
g^1_{12} = -{{3\,u^1}\over{(p^2_2-u^1u^2)^{2/3}}};\\
\fl
d \sigma^2_2=\left(-g^1_{22} + {{2\,(-2\,(p^2_2)^2+7\,u^1u^2p^2_2
-5\,(u^1u^2)^2+2\,(u^1)^3-3\,u^1\,p^1_2)}\over{(p^2_2-u^1u^2)^2}}\right)\,
\xi^1 \wedge \xi_2+...\\
~~\Rightarrow~~
g^1_{22} = {{2\,(-2\,(p^2_2)^2+7\,u^1u^2p^2_2-5\,(u^1u^2)^2+2\,(u^1)^3
-3\,u^1\,p^1_2)}\over{(p^2_2-u^1u^2)^2}}.
\end{eqnarray*}
Thus all the group parameters are expressed as the functions of the local
coordinates $\{x^1,x^2,u^1,u^2,p^1_2,p^2_2\}$ of the equation
${\mbox{{\Rc}}}_1$. The result of all normalizations is the invariant coframe
\begin{eqnarray*}
\fl \theta^1 = {{d u^1-(p^2_2-u^1u^2)\,d x^1 - p^1_2\,d x^2}
\over{(p^2_2-u^1u^2)^{2/3}}},~~~
\\
\fl
\theta^2 = {{d u^2-u^1\,d x^1 - p^2_2\,d x^2}\over{(p^2_2-u^1u^2)^{1/3}}},\\
\fl \xi^1 = (p^2_2-u^1u^2)^{1/3}\,(d x^1+u^2\,d x^2),~~~
\\
\fl
\xi^2 = (p^2_2-u^1u^2)^{2/3}\,d x^2,\\
\fl \sigma^1_2 = {{d p^1_2-u^2 dp^2_2+((u^2)^2-2\,u^1)\,du^1
+u^1u^2\,du_2}\over{(p^2_2-u^1u^2)^{4/3}}}\\
\fl ~~~~+{{u^1(p^2_2-u^1u^2)\,d x^1+(4\,(u^1)^3-7\,(u^1u^2)^2
+11\,u^1u^2p^2_2-4\,u^1p^1_2-4\,(p^2_2)^2)\,dx_2}
\over{(p^2_2-u^1u^2)^{4/3}}},\\
\fl \sigma^2_2 = {{d p^2_2-u^2 du^1-u^1 du^2-(p^1_2+u^1(u^2)^2
-(u^1)^2-u^2p^2_2)\,d x^1}\over{p^2_2-u^1u^2}}\\
\fl ~~~~ +{{(4\,(u^1)^2u^2+(u^2)^2p^2_2-u^1(u^2)^3-u^2p-3\,u^1p^2_2)\,d x^2}
\over{p^2_2-u^1u^2}}.
\end{eqnarray*}
\noindent
Its structure equations are
\begin{eqnarray*}
d\theta^1 = I\,\theta^1\wedge \xi^1
+\case23\,\theta^1\wedge\sigma^2_2
+\xi^1\wedge\sigma^2_2
+\xi^2\wedge\sigma^1_2,\\
d\theta^2 = -\theta^1\wedge \xi^1
+\case12\,I\,\theta^2\wedge\xi^1
+\case13\,\theta^2\wedge\sigma^2_2
+\xi^2\wedge\sigma^2_2,\\
d\xi^1 = \theta^2\wedge \xi^2
-\case13\,\xi^1\wedge\sigma^2_2,\\
d\xi^2 = I\,\xi^1\wedge \xi^2
-\case23\,\xi^2\wedge\sigma^2_2,\\
d\sigma^1_2 = -\theta^1\wedge \xi^1
-6\,I\,\theta^1\wedge\xi^2
-\case32\,I\,\theta^2\wedge\xi^1
-\theta^2\wedge\sigma^2_2
-15\,I\,\xi^1\wedge\xi^2\\
~~~~-2\,I\,\xi^1\wedge\sigma^1_2
+7\,\xi^2\wedge\sigma^2_2
+\case43\,\sigma^1_2\wedge\sigma^2_2,\\
d\sigma^2_2 = -3\,\theta^1\wedge \xi^2
+\theta^2\wedge\xi^1
-\case32\,I\,\theta^2\wedge\xi^2
+\xi^1\wedge\sigma^1_2
-\case32\,I\,\xi^1\wedge\sigma^2_2,
\end{eqnarray*}
\noindent
where the only invariant $I$ is of form
\[
I={{2\,(p^1_2+u^1(u^2)^2-(u^1)^2-u^2p^2_2)}\over{3\,(p^2_2-u^1u^2)^{4/3}}}.
\]
\noindent
Taking its exterior differential, we have
\[
d I = -\case23\,\theta^2 - 2\,I^2\,\xi^1+2\,\xi^2
+\case23\,\sigma^1_2-\case43\,I\,\sigma^2_2,
\]
\noindent
so all differential invariants of the group are functionally expressed
as functions of $I$, the rank of the coframe \cite[Proposition 8.18]{Olver95}
is equal to 1, and its symmetry group is 5-dimensional
\cite[Theorem 8.22]{Olver95} (as it should be;
for the full details of finding infinitesimal generators of this group
by Lie's method see, e.g.,  \cite[Chapter 3, {\parsign}  5]{VK99}.)

\subsection{Example 2: One-dimensional equations of gas dynamics in Lagrange
coordinates}

One-dimensional dynamics of polytropic gas in Lagrange coordinates is
described by the system of d.e.s \cite{RYa}
\begin{eqnarray}
\rho_t+\rho^2 u_m=0,\nonumber\\
u_t+p_m=0,\label{gas}\\
p_t+\gamma\,\rho\,p\,u_m=0.\nonumber
\end{eqnarray}
\noindent
Denoting $\rho=u^1$, $u=u^2$, $p=u^3$, $t=x^1$, $m=x^2$ and using the
above method, we obtain the invariant coframe of the symmetry group of the
system (\ref{gas})
\begin{eqnarray}
\fl \theta^1 = {1\over{u^1}}\,\left(d u^1
+(u^1)^2 p^2_2\,d x^1 - p^1_2\, d x^2  \right), ~~~
\nonumber\\
\fl
\theta^2 = \sqrt{{{u^1}\over{\gamma u^3}}}\,\left(d u^2
+p^3_2\,d x^1- p^2_2\, d x^2\right),\nonumber\\
\fl \theta^3 = {1\over{\gamma u^3}}\,\left(d u^3
+\gamma\,u^1\,u^3\,p^2_2\,d x^1- p^3_2\, d x^2\right),
\nonumber\\
\fl
\xi^1 = \sqrt{{{u^1}\over{\gamma u^3}}}\,d x^2,
\label{gas_coframe}\\
\fl
\xi^2 = u^1\,p^2_2\,d x^1,\nonumber\\
\fl \sigma^1_2 = {1\over{u^1\,p^2_2}}\,\sqrt{{{\gamma u^3}\over{u^1}}}\,
\left(d p^1_2 - {{p^1_2} \over {u^1}}\,d u^1 -{{(\gamma-1)\,(u^1)^3
(p^2_2)^2u^3-(p^1_2)^2(u^3)^2-(p^3_2)^2(u^1)^2}\over{2\,u^1\,(u^3)^2}}\,d x^2
\right),\nonumber\\
\fl \sigma^2_2 = {1\over{u^1\,p^2_2}}\,\left(d p^2_2
+{{\gamma-1}\over 2}\,(u^1)^2(p^2_2)^2\,dx^1 +p^1_2\,p^2_2\,d x^2\right),
\nonumber\\
\fl \sigma^3_2 = {{1}\over{p^2_2\,\sqrt{\gamma\,u^1\,u^3}}}\,
\left(d p^3_2
+\gamma\,u^1\,p^2_2\,p^3_2\,d x^1- {{\gamma-1}\over{2}}\,u^1\,(p^2_2)^2\,
d x^2\right)
\nonumber
\end{eqnarray}
\noindent
(since from considering the physical meaning we have $u^1=\rho>0$  and
$u^3 = p>0$, therefore there is no need to worry about the signs of the
expressions under the square roots).

The structure equations of this coframe are
\begin{eqnarray*}
d \theta^1 = \theta^1\wedge \xi^2
    +\xi^1\wedge\sigma^1_2-\xi^2\wedge \sigma^2_2,\\
d \theta^2 = \case12\,\theta^1\wedge \theta^2
    +\case{\gamma}2\,\theta^2\wedge\theta^3
    +I_1\,\theta^2\wedge\xi^1
    +\case{\gamma-1}2\,\theta^2\wedge\xi^2
    +\xi^1\wedge\sigma^2_2
    -\xi^2\wedge \sigma^3_2,\\
d \theta^3 = \theta^1\wedge\xi^2
   + I_2\,\theta^3\wedge \xi^1+\xi^1\wedge\sigma^3_2
   -\xi^2\wedge\sigma^2_2,\\
d \xi^1 = \case12 \theta^1\wedge \xi^1
   -\case{\gamma}2\theta^3\wedge\xi^1
   -\xi^1\wedge\sigma^2_2,\\
d \xi^2 = \theta^1\wedge \xi^2
   -\xi^2\wedge \sigma^2_2,\\
d\sigma^1_2 = \case12\,\gamma\,(\gamma-1)\,\theta^1\wedge\xi^1
   -\case12 \,\theta^1\wedge\sigma^1_2
  -\case12\,\left(2\,I^2_2-\gamma^2+\gamma\right)\,\theta^3\wedge\xi^1\\
~~~+\case{\gamma}2\,\theta^3\wedge\sigma^1_2
   +I_1\,\xi^1\wedge\sigma^1_2
   +\gamma\,(\gamma-1)\,\xi^1\wedge\sigma^2_2
   -\gamma\,I_2\,\xi^1\wedge\sigma^3_2
   +\sigma^1_2\wedge\sigma^2_2,\\
d\sigma^2_2=\case{\gamma-1}2\,\theta^1\wedge\xi^2
   -\xi^1\wedge\sigma^1_2
   -\case{\gamma-1}2\,\xi^2\wedge\sigma^2_2,\\
d\sigma^3_2 = -\case{\gamma-1}2\,\theta^1\wedge\xi^1
   +I_2\,\theta^1\wedge\xi^2
   -\case12\,\theta^1\wedge\sigma^3_2
   -\case{\gamma}2\,\theta^3\wedge\sigma^3_2
   +(\gamma-1)\,\xi^1\wedge\sigma^2_2\\
~~~-I_1\,\xi^1\wedge\sigma^3_2
   -I_2\,\xi^2\wedge\sigma^2_2
   -\sigma^2_2\wedge\sigma^3_2.
\end{eqnarray*}
\noindent
The invariants $I_1$ and $I_2$ are defined by the equalities
\[
I_1 = \sqrt{{{\gamma\,u^1}\over{u^3}}}\,{{p^3_2\,u^1 - p^1_2\,u^3}\over
{2\,(u^1)^2\,p^2_2}},~~~~
I_2=\sqrt{{{\gamma}\over{u^1\,u^3}}}\,{{p^3_2}\over{p^2_2}}.
\]
\noindent
Their exterior differentials are
\begin{eqnarray*}
d I_1 = -{{I_1}\over{2}}\,\theta^1 +{{\gamma}\over 2}\,(I_1-I_2)\,
\theta^3+ {1\over 2}\,\sigma^1_2-I_1\,\sigma^2_2
+{\gamma\over 2}\,\sigma^3_2,\\
d I_2 =-{{I_2}\over 2}\,\theta^1+\gamma\,\left(I_1-{{I_2}\over 2}\right)
\,\theta^2+\left({{\gamma\,(\gamma-1)}\over{2}}-I_1I_2\right)\,
\xi^1-I_2\,\sigma^1_2+\gamma\,\sigma^3_2,
\end{eqnarray*}
\noindent
so all differential invariants of the symmetry group depend functionally on
$I_1$ and $I_2$. Thus the coframe (\ref{gas_coframe}) has the rank 2, and
the symmetry group of the system (\ref{gas}) is 6-dimensional.
In \cite[Chapter 3]{AGI} the explicit form of the infinitesimal generators
of this group is given.

\subsection{Example 3: Liouville's equation}

For finding invariant 1-forms and structure equations of the symmetry
pseudo-group of Liouville's equation
\[
u_{tx} = \e^u
\]
\noindent
we take the equivalent system of the first order
\[
u_t = v, ~~~~~ v_x = \e^u.
\]
\noindent Using the notations $u=u^1$, $v = u^2$, $t=x^1$, $x = x^2$ and
applying the above procedure of absorption and normalization, we have
$\sigma^1_1 = 0$, $\sigma^2_2 = 0$, while $\theta^1$, $\theta^2$, $\xi^1$,
$\xi^2$, $\sigma^1_2$ and $\sigma^2_1$ constitute the lifted coframe
\begin{eqnarray}
\theta^1 = d u^1 - u^2 d x^1 - p^1_2 d x^2,\nonumber\\
\theta^2 = a^2_2 \left(d u^2 - p^2_1 d x^1 - \e^{u^1} dx^2 \right),\nonumber\\
\xi^1 = (a^2_2)^{-1} dx^1, \label{L_cf_1}\\
\xi^2 = a^2_2 \e^{u^1} \d x^2, \nonumber\\
\sigma^1_2 = (a^2_2)^{-1} \e^{-u^1} d p^1_2 - (a^2_2)^{-1} d x^1
+ a^2_2 g^1_{22} \e^{u^1} d x^2, \nonumber\\
\sigma^2_1 = (a^2_2)^2 \left(d p^2_1 - u^2 d x^1 + ((a^2_2)^{-1} g^2_{11}
+ u^2 p^2_1) d x^1 \right). \nonumber
\end{eqnarray}
The exterior differentials of these forms are
\begin{eqnarray}
d \theta^1 = - \theta^2\wedge \xi^1 + \xi^1 \wedge \sigma^1_2,\nonumber\\
d \theta^2 = \chi_1 \wedge \theta^2 - \theta^1\wedge \xi^2
+ \xi^1\wedge\sigma^2_1,\nonumber\\
d \xi^1 = - \chi_1 \wedge \xi^1,\label{L_se_1}\\
d \xi^2 = \chi_1\wedge \xi^2 + \theta^1\wedge \xi^2,\nonumber\\
d \sigma^1_2 = \chi_2\wedge \xi^2 - \chi_1 \wedge \sigma^1_2
-\theta^1 \wedge (\sigma^1_2 + \xi^1), \nonumber\\
d \sigma^2_1 = \chi_3 \wedge \xi^1 + 2\,\chi_1\wedge \sigma^2_1,
\nonumber
\end{eqnarray}
\noindent
where
\begin{eqnarray}
\chi_1 = (a^2_2)^{-1} d a^2_2 + a^2_2 u^2 \xi^1, \nonumber\\
\chi_2 = d g^1_{22} + 2\,g^1_{22} (\chi_1+\theta^1)
+(a^2_2)^{-1} \e^{-u^1} p^1_2 (\xi^1-\sigma^1_2)+w_1 \xi^2, \label{L_cf_2}\\
\chi_3 = d g^2_{11} - 3\,g^2_{11} \chi_1
+ (a^2_2)^2 (p^2_1+(u^2)^2)\,(\theta^2+\xi^2)+3\,a^2_2u^2 \sigma^2_1
+ w_2 \xi^1, \nonumber
\end{eqnarray}
\noindent
$w_1$ and $w_2$ are free parameters. The structure equations (\ref{L_se_1})
do not contain any torsion coefficient depending on the group parameters.
The coframe (\ref{L_cf_1}) is not involutive, because its degree of
indeterminancy $r^{(1)}$ is 2, whereas the reduced characters are
$s^{\prime}_1 = 3$, $s^{\prime}_2 = ... = s^{\prime}_6 = 0$, so Cartan's test
is not satisfied. Therefore we should use the procedure of prolongation
\cite[Chapter 12]{Olver95}. For this purpose we unite both coframes
(\ref{L_cf_1}) and (\ref{L_cf_2}) into the new base coframe, whereas
$w_1$ and $w_2$ turn into the new group parameters. Finding exterior
differentials of $\chi_1$, $\chi_2$ and $\chi_3$, we have
\begin{eqnarray}
d \chi_1 = \theta^2 \wedge \xi^1 - \xi^1 \wedge \xi^2, \nonumber\\
d \chi_2 = \nu_1 \wedge \xi^2 - 2\,\theta^1 \wedge \chi_1
- 2\, \chi_1 \wedge \chi_2, \label{L_se_2}\\
d \chi_3 = \nu_2 \wedge \xi^1 + 2\, (\theta^2+\xi^2)\wedge\sigma^2_1
+3\, \chi_1\wedge\chi_2,\nonumber
\end{eqnarray}
\noindent
where
\begin{eqnarray*}
\fl
\nu_1 = d w_1 + 3\,w_1 (\theta^1+\chi_1)
+ \left((a^2_2)^{-1} \e^{-2 u^1} (p^1_2)^2 - g^1_{22} \right)\,
(\xi^1+\sigma^1_2) -(a^2_2)^{-1}\e^{-u^1} p^1_2 \chi_2,\\
\fl
\nu_2 = dw_2 + 4\,w_2 \chi_2 + 2\,\left((a^2_2)^3(u^2)^3
- g^2_{11}\right)\,(\theta^2+\xi^2)
+ 2\,(a^2_2)^2\left((u^2)^2-2\,p^2_1\right)\, \sigma^2_1\\
+ 3\,a^2_2 u^2 \chi_3.
\end{eqnarray*}
The structure equations (\ref{L_se_2}) admit the change
\[
\nu_1  \mapsto  \nu_1 + z_1 \xi^2, ~~~~
\nu_2  \mapsto  \nu_2 + z_2 \xi^1
\]
\noindent
for the free parameters $z_1$ and $z_2$. So the degree of indeterminancy of
the coframe (\ref{L_cf_1}), (\ref{L_cf_2}) is $r^{(1)} = 2$ again,
while the reduced characters now are $s^{\prime}_1 = 2$,
$s^{\prime}_2 = ... = s^{\prime}_9 = 0$.  Cartan's test is therefore
satisfied, and the coframe (\ref{L_cf_1}), (\ref{L_cf_2}) is involutive.
Since the last non-zero reduced character is $s^{\prime}_1 = 2$,
the symmetry pseudo-group transformations depend on two arbitrary functions of
one variable. This agrees with the result found by Liouville \cite{Liouville}.
In \cite{LisleReidBoulton,LisleReid} the structure equations of this
pseudo-group are derived using a different method; see also
\cite{ReidBoultonLisle}.

\section{Conclusion}

The approach to computation of symmetry groups used here does not require
obtaining infinitesimal defining systems, analysis of their involutivity
and integration, and includes only differentiation and linear algebra
operations. So it is algorithmic {\it in principle}, although the labyrinth
of corresponding computations is very intricate. In the future it seems that
it will be possible to reduce the complexity of computations by means of using
the canonical contact forms \cite{Olver2000} on bundles of higher order jets.

\end{document}